# Programmable Electromagnetic Space via Metasurface Clusters

Min Li[1], Lixiang Meng[1], Gongxu Dong[1], Xiaobo Zhou[1], Lu Song[2, *], Puti Yan[3, *], Dashuang Liao[4], Chao Qian[5], Zuojia Wang[5], and Hongsheng Chen[5, *]

[1] College of Electronics and Electric Engineering, Anhui Agricultural University, Hefei, 210036, China.

[2] Air and Missile Defense College, Air Force Engineering University, Xi' an 710051, China.

[3] College of Electronic Information Engineering, Beihang University, Beijing, 100191, China

[4] Department of Computer Science, Anhui Medical University, Hefei, 230032, China.

[5] ZJU-UIUC Institute, Interdisciplinary Center for Quantum Information, State Key Laboratory of Extreme Photonics and Instrumentation, Zhejiang University, Hangzhou 310027, China.

*Corresponding author: mlsonglu@163.com (L. Song); Classicypt1230@163.com (P. Yan); hansomchen@zju.edu.cn (H. Chen);

**Abstract:** The rapid evolution of next-generation communications and the Internet of Things (IoT) has catalyzed an urgent demand for governing expansive spatial environments as functional electromagnetic (EM) entities. However, deterministically programming such open EM spaces remains a formidable challenge, as current methodologies are largely confined to localized interfaces that lack the collective coordination required to orchestrate unbounded environments. Here, we introduce a general framework for the deterministic programming of EM space via cooperative metasurface clusters, achieved by mapping volumetric field interference landscapes onto a virtual nodal network. By representing excitations and meta-atoms as fully interconnected nodes, we transform intricate non-local interactions into tractable nodal states, enabling the precise quantitative synthesis of spatial scattering. This framework bridges local meta-atoms with global EM environment to program space as a functional entity, as demonstrated by a deeply coupled meta-emitter for programmable collective radiation and metasurface clusters that sculpt angle-resolved illusion spaces. By transitioning from individual components to cooperative multi-body assemblies, our work provides a scalable foundation for next-generation wireless networks, wave-based analog computing, and ambient intelligence, where space itself becomes a coherent functional and reconfigurable entity capable of holistic information management.

**Introduction**

The rapid evolution of sixth-generation (6G) communications, the Internet of Things (IoT), and intelligent systems is fundamentally reshaping our interaction with the physical world, driving an urgent need to transform expansive environments into programmable electromagnetic (EM) entities [1,2]. Realizing this vision of ambient intelligence requires the ability to orchestrate not only individual devices but also the EM properties of space itself [3,4]. Consequently, characterizing and manipulating EM waves in unbounded environments have emerged as a cornerstone of modern wave-based wireless technologies [5]. In such environments, wave propagation is fundamentally influenced by multiple scattering [6-8] and non-local interactions [9-11] among distributed scatterers, collectively forming an EM space characterized by the global scattering landscape [12]. Decades ago, the forays into metasurfaces have marked a transformative step in EM wave control through precise engineering of local boundary conditions [13-15]. However, these artificially structured interfaces are typically localized and treated as isolated elements embedded in space [13,16], which limits their capacity to govern the non-local wave dynamics inherent in expansive environments. This constraint necessitates a transition from boundary-layer intervention toward cooperative metasurface clusters, where collective interactions among distributed assemblies emerge as a new degree of freedom for shaping wave scattering at the space level.

Motivated by this perspective, increasing attention has turned to multi-body systems of interacting scatterers or metasurfaces [17-27] to harness collective wave scattering. Strong inter-element coupling and multiple scattering give rise to complex interaction spaces that serve as physical resources [17,28,29]. Recent experimental studies have explored these platforms for diverse wave-based functionalities, including neuromorphic computing implemented in multiple-scattering media [17], signal differentiation processing in closed chaotic cavities [18,19], and nonlinear computation enabled by high-dimensional scattering dynamics in linear media [20]. To further expand and diversify the available EM space, distributed metasurface architectures have been developed. For instance, spatially distributed and independent metasurfaces have demonstrated the capacity for intelligent sensing and multi-scale wave control [21]. Complementarily, metasurface clusters exploit collective interactions and joint scattering among multiple units to achieve non-contact stealth by transforming individual scatterers into a cooperative spatial entity [22]. Despite these advances, current methodologies largely rely on empirical calibration or data-driven optimization [17-22]. In the absence of a physically transparent model, these approaches typically construct surrogate models to approximate the input–output mapping of the system. Only a limited portion of the available scattering information is effectively utilized, leaving much of the high-dimensional scattering space unexplored. This reliance on empirical optimization stems from the lack of a comprehensive theoretical framework capable of predicting and controlling the complex multi-body coupling and non-local interactions that govern collective scattering in distributed scatterers. Consequently, achieving programmable and task-specific control of the EM space in unbounded environments remains an open challenge, hindering both the fundamental understanding of complex wave physics and principled synthesis of space-level functionalities.

In this work, we introduce a theoretical framework that that renders EM space formed by distributed metasurface clusters fully programmable. The central idea is to map complex EM field interactions onto a virtual nodal network where excitations and meta-atoms are represented as fully interconnected nodes. This representation converts otherwise intractable non-local field interactions into tractable nodal variables. The evolution of these nodal states captures the collective wave dynamics and allows deterministic computation and synthesis of the resulting global scattering. Building on this framework, cooperative metasurface assemblies emerge as programmable platforms for space-level EM functionalities, as demonstrated by a deeply coupled meta-emitter for programmable radiation and metasurface clusters that cooperatively sculpt an angle-resolved illusion space. Our work represents a paradigm shift from individual metasurface to cooperative ensembles, enabling deterministic control over complex wave scattering within open media. More broadly, it reimagines the EM environment as distributed computational fabric, unlocking transformative potential for wave-based computing, high-capacity information processing that harnesses the full spatial scattering landscape, and the holistic orchestration of space with bespoke functionalities.

## Results

**Theoretical framework**

The accurate control of EM responses in structurally complex environments has long been impeded by local approximation limits. Traditional local paradigms treat metasurfaces as collections of isolated elements, each assumed to operate independently. However, this localized view breaks down in disordered metasurface clusters with arbitrary topologies (Fig. 1a). In such multi-body systems, implementation of space-level EM modulation becomes notoriously difficult, as wave propagation is governed by strong inter-element coupling and complex multiple scattering (left panel in Fig. 1b). These collective interactions create a non-local dependency where the state of a single meta-atom influences the global field distribution, rendering conventional interface-level programming insufficient for capturing the full EM landscape.

To resolve these challenges, we introduce the paradigm of spatial programming based on virtual nodal network, shifting the focus from boundary-layer intervention to the holistic orchestration of EM fields across spatial volumes. In this representation, the complex EM field landscape in physical space is mapped onto a fully interconnected network of coupled nodes (Fig. 1b). Within the network, the specific geometry and material properties of each element are abstracted into discrete, coupled nodes. Crucially, the network formalization categorizes these nodes into two distinct functional groups based on their roles as primary or secondary sources. The excitation group encompasses all primary sources, including traditional antenna arrays or active radiating metasurfaces [30,31] that initiate energy injection. In contrast, the metasurface group comprises the remaining passive meta-atoms that contribute to the secondary scattering process [32]. It is important to emphasize the fully interconnected nature of this framework, i.e., every element is coupled not only to the excitations but also to every other element within the clusters. This all-to-all connectivity ensures that the network captures the full complexity of mutual coupling, allowing the system to be solved as a unified entity.

The physical interpretability of the coupling network is established by assigning a nodal voltage and an intrinsic radiation pattern to each node. The excitation nodes represent physical feeding ports of the system, and their nodal voltages correspond to the applied port voltages of the excitation sources. In contrast, to extract a network representation of the metasurface cluster, a deeply subwavelength region associated with each meta-atom is designated as a virtual lumped port (see

Supplementary Note 1). By characterizing these ports through a global mutual impedance or scattering matrix (Supplementary Note 2), the framework rigorously encapsulates all-to-all mutual coupling and multiple scattering properties. We further derived the analytic expression for the complete nodal voltage vector as a coherent superposition of individual excitations, which is written as (Supplementary Note 2):

$$\boldsymbol{V} = \sum_i \boldsymbol{V}_i \cdot \frac{V_E^i}{V_i(i)}, \boldsymbol{V}_i = \boldsymbol{\mathcal{F}}_i \cdot \boldsymbol{V}_E^i \qquad (1)$$

Here, $\boldsymbol{Z}_M$ denotes the impedance distributions of virtual meta-atom ports, $\boldsymbol{V}_E^i$ is defined as excitation vector $[0, …,V_E^i, …, 0]^T$ where the i-th element is the scalar $V_E^i$ and all other entries are zero, $V_i(i)$ refers to the i-th element of the array $V_i$. $\boldsymbol{\mathcal{F}}_i$ is a function of the meta-atom impedance distribution. Eq. (1) indicates that the voltages at the virtual ports of the meta-atoms are jointly determined by the external excitation and the local impedance distribution of all meta-atoms. This reflects the strong consistency of the proposed framework with the underlying physical scattering process: on the one hand, it captures the mutual coupling among distributed elements; on the other hand, it reveals that each meta-atom acts as a secondary radiator excited by primary sources.

The intrinsic radiation pattern provides the fundamental wave information carrier within the disordered environment (Supplementary Note 3). The global far-field is then synthesized by a linear superposition of these intrinsic patterns weighted by the normalized node voltages:

$$\boldsymbol{f}_{total}(\theta, \varphi) = \sum_i \frac{V(i)}{\sqrt{\sum_j |V(j)|^2}} \cdot \boldsymbol{f}_i(\theta, \varphi) \qquad (2)$$

Here V(i) and V(j) are the i-th and j-th node voltages calculated from Eq. (1), $\boldsymbol{f}_i(\theta, \varphi)$ refers to the intrinsic radiation field of i-th node. The linkage between network nodal states and physical EM scattering is explicitly defined by Eq. (2). It establishes a general framework for the efficient modeling and calculation of the global scattering of highly coupled and randomly distributed metasurfaces. For information intelligent metasurfaces [33], local modulations of meta-atom impedance $\boldsymbol{Z}_M$ via switching PIN diodes trigger an instantaneous and predictable redistribution of nodal states across the entire network. By resolving this global redistribution through a unified matrix equation, the proposed theory bypasses the computational bottleneck of iterative full-wave simulations, enabling the rapid synthesis of complex radiation fields. Consequently, by inversely designing the distribution of meta-atom impedance distributions, the EM space formed by metasurface clusters is transformed into a fully programmable entity.

To quantitatively validate the proposed theory in a realistic setting, we consider a representative configuration comprising two dipole antennas as excitation sources and two metasurface meta-atoms, as illustrated in Fig. 2a. For illustrative purposes, a square ring resonator is adopted as the representative meta-atom. The proposed framework, however, is general and applies to arbitrary meta-atom geometries, including more complex configurations discussed later in this work. Detailed parameters of this representative scene are provided in Supplementary Note 4. A deeply subwavelength region (red area) is modeled as a virtual lumped port characterized by a nodal voltage and an impedance model. As a result, the system is represented by a four-node coupling network comprising two excitation nodes and two meta-nodes, which are sequentially indexed as nodes 1–4. By tuning the impedance states of the meta-atoms, a set of representative port impedances is obtained and distributed on the complex plane, namely 500+22.7jΩ, -190-5jΩ, 0+0jΩ and 0+∞jΩ as shown in Fig. 2b. These impedance values correspond to distinct physical boundary conditions at the effective ports. An impedance of zero represents a short-circuit condition, under which the meta-atom behaves as a closed metallic loop, whereas an infinite impedance corresponds to an open-circuit condition, associated with an open split-ring resonator. The sign of the imaginary part of the impedance reflects the presence of loss or gain, indicating meta-atoms incorporating dissipative or active elements, respectively. For each impedance configuration, the excitation-node voltages are fixed as $V_E^1 = 1e^{j\pi/2}$, $V_E^2 = 1$, and the corresponding meta-node voltages are analytically derived shown in Fig. 2c. To provide a complete map of the mutual coupling effects across the entire meta-system, the global scattering matrix is extracted as shown in Fig. 2d. The theoretical three-dimensional far-field scattering patterns, derived from the resulting nodal voltages, align perfectly with full-wave simulations (Fig. 2e), a consistency further validated by the normalized two-dimensional profiles in Fig. 2f. This accuracy extends to key metrics such as directivity, confirming that the framework precisely captures the angular scattering characteristics. However, a slight deviation occurs in the realized gain, where theoretical predictions appear marginally lower in certain cases (Fig. 2g). This discrepancy is primarily rooted in the combined effects of return loss and radiation leakage inherent to the auxiliary ports within the nodal network. Specifically, suboptimal impedance realizations trigger enhanced power reflections via impedance mismatch, while simultaneously diminishing the coupling efficiency between internal modes and free-space radiation. As a result, the total radiated power is suppressed without distorting far-field scattering pattern, leading to a reduction in realized gain while leaving the intrinsic directivity intact. Despite this discrepancy, the network framework remains a robust and high-fidelity approach for predicting global scattering behavior. By explicitly encoding intricate non-local interactions into a systematic network representation, it enables deterministic wave-field synthesis in metasurface clusters beyond the reach of conventional design methodologies.

**Programmable collective radiation via a deeply coupled meta-emitter**. To demonstrate the spatial programming capabilities of our framework, we first investigate the synthesis of programmable collective radiation within a deeply coupled meta-system (Fig. 3a, b). In conventional metasurface designs, inter-element coupling is typically treated as a parasitic effect to be suppressed. Wave manipulation thus relies on the local phase engineering of isolated and uncoupled meta-atoms (Fig. 3a). In contrast, our framework treats excitations and metasurfaces as a strongly coupled many-body system, where inter-element interactions are harnessed to collectively determine the radiation profile (Fig. 3b). The hallmark of this collective control is the joint optimization of global degrees of freedom. Specifically, the impedance states of the meta-atoms modulated via PIN diodes, are optimized in tandem with the excitation phases of the feeding network. This synergistic approach allows us to sculpt the EM environment with unprecedented precision, mapping desired far-field objectives back to the required nodal states through our deterministic framework.

To validate this approach, we implemented a hybrid meta-emitter comprising a patch antenna array and a transmissive metasurface featuring 1-bit discrete impedance states as the modulation layer shown in Fig. 3b. Detailed geometric parameters and performance characterization are provided in Supplementary Note 5. The versatility of this joint optimization method is subsequently demonstrated through the synthesis of multifunctional radiation patterns, ranging from single-beam to multi-beam configurations (Fig. 3c–e). The optimization algorithms are included in **Methods**. For each scenario, the optimized metasurface topologies and their corresponding excitation phase distributions are presented (left and middle panels). The resulting radiation patterns (right panels) show that the theoretical predictions are in excellent agreement with full-wave simulations, accurately capturing the beam positions even the side-lobe levels and null depths. Notably, the optimized excitation profiles exhibit a quasi-random, non-intuitive distribution that deviates significantly from conventional spherical or planar wavefronts. By liberating the system from the constraints of fixed excitation, our framework elevates the source distribution into an active, programmable degree of freedom. This co-design method expands the synthesis space into a higher dimension, where the coordinated states of feed phases and meta-atoms collectively encode the spatial field. Such a capability is particularly promising for advanced beamforming [34] and physical-layer secure communications [35], as the global EM field can only be accurately reconstructed through the precise synchronization of both excitations and metasurface configurations.

**Programmable angle-resolved illusion space**. To further explore the potential of the virtual nodal network in orchestrating complex spatial landscapes, we demonstrate the synthesis of a programmable, angle-resolved illusion space. Unlike static illusions, this concept leverages a spatially distributed metasurface cluster to produce angle-dependent EM signatures as shown in Fig. 4a. By coordinating the cooperative scattering of multiple metasurfaces illuminated by a multi-beam transmitter array (Fig. S8), the EM environment is programmed to project distinct scattering landscapes. This effectively creates a volumetric mirage, where a remote receiver detects the scattering profile of a specific, pre-defined virtual object (Object 1–3). The metasurface panels are composed of 1-bit reconfigurable reflective meta-atoms, as shown in Fig. 4b. Detailed information on the spatial arrangement of the metasurfaces, as well as the geometry and EM responses of the meta-atoms, is provided in Supplementary Note 6. The incident wave is generated by an array of eight patch antennas, as depicted in the top panel of Fig. 4c. By programming the relative phase delays among the antennas, the direction of the incident wave can be steered. Here, three representative incident angles of 0°, 20°, and -20° are considered, with the corresponding radiation patterns shown in Fig. 4c and labelled as beam #1, #2, and #3. As the transmitter array generates multiple excitation beams with diverse radiation directions, it triggers distinct collective scattering responses across the global space. Through inverse spatial programming, we concurrently optimize the meta-atom states to reconstruct the target scattering fingerprints of three disparate objects in three separate views (Figs. 4d–f). The theoretical predictions (scatter points) show remarkable agreement with full-wave simulations (light solid curves), both of which faithfully replicate the target scattering profiles (dark solid curves). The insets indicating the programmed state distributions of the three metasurface panels, where the arrow in each coding pattern specifies the corresponding coding orientations and are consistent with those marked in Fig. 4a. The spatial configuration of metasurface clusters and geometries of target objects are included in Fig. S11 and S12. Details of the optimization algorithm are included in **Methods**. We further quantify the reconstruction similarity [36] under both activated and deactivated metasurface conditions (Figs. 4g-i). When the metasurface cluster is activated, the similarity between the synthesized and target scattering distributions remains consistently high across all perspective-dependent views. In contrast, deactivating the metasurfaces (all meta-atom switched on identical state) leads to a dramatic collapse in similarity, reducing the environment to its disordered baseline. Notably, in View 1, the reconstruction similarity remains relatively high even when the metasurface cluster is deactivated. This phenomenon occurs because the primary radiation direction of the incident beam aligns closely with our observed scattering plane, allowing the incident wave to dominate the far-field response (see Fig. S15-17) In contrast, for the other two cases, the excitation beams are steered away from the observation planes. Consequently, the passive environment fails to replicate the target profiles and the active modulation of the metasurface cluster becomes the decisive factor, leading to a substantial enhancement in similarity. The results underscore a critical physical insight where the spatial positioning of metasurfaces profoundly dictates their control authority over the EM environment. Optimizing the global placement of meta-clusters remains beyond the scope of this work, yet it emerges as a compelling frontier for future research.

**Experimental demonstration of programmable illusion space via metasurface clusters.**

To experimentally validate the programmable illusion capabilities of the synergistic metasurface clusters under varying illuminated angles, we fabricated the metasurface samples and conducted experimental characterization in a microwave anechoic chamber. As depicted in Fig. 5a, a transmitter array produces multiple excitation beams with distinct illumination directions, effectively interrogating the EM space formed by the distributed metasurface clusters. Detailed fabrication procedures and measurement setups are provided in the **Methods** section. The measured radiation characteristics confirm the generation of three directional beams #1-#3 as shown in Fig. 5b. To ensure the reconfigurability of the metasurface, we first characterize its reflection amplitude and phase under different PIN diode switching states (Fig. S13). The results shown in Fig. 5c and 5d indicate that, upon switching the PIN diode states, the metasurface maintains a consistently high reflection amplitude, while a nearly 180° phase shift is achieved at the operating frequency of 10 GHz. Next, we measure the radar cross-section (RCS) of three target objects and compare them with the RCS reconstructed by the metasurface under the guidance of the virtual nodal network. The reconstructed far-field scattering under varying illumination directions are illustrated in Fig. 5e–g. The experimental comparisons demonstrate that the scattering patterns generated by the metasurface clusters closely reproduce those of the corresponding reference objects across all three illumination configurations, which further validates the universality of the proposed methodology. Minor deviations in the peak directions are observed between the measured scattered fields of the reference objects and the metasurface clusters. These discrepancies are primarily attributed to mechanical alignment errors of the reference objects during measurement and parasitic scattering from the DC biasing cables.

**Discussion**

In summary, we have established a deterministic framework for programming EM space by harnessing the collective scattering of metasurface clusters. Departing from the conventional focus on localized, isolated interfaces, our approach treats the entire environment as a fully interconnected nodal network. This transition allows us to resolve intricate non-local field interactions, traditionally viewed as parasitic interference, and transform them into exploitable degrees of freedom. By rendering global nodal states tractable, we enable the sculpting of complex field landscapes highly precisely.

Compared to empirical or data-driven models, this nodal representation offers superior physical transparency and scalability. It provides a direct predictive link between local meta-atomic states and the global scattering, precluding the need for exhaustive retraining in new environments. Furthermore, this framework paves the way for distributed analog computing [37] and integrated sensing-communication-computing (ISCC) systems [38]. By viewing metasurface clusters as a physical neural network, the environment itself becomes a volumetric processor [5], where collective scattering mimics synaptic weights to perform high-dimensional information processing at the speed of light.

Future research should transition from position-fixed programming to autonomous spatial optimization. While we have mastered the control of nodal states, the joint optimization of nodal states and physical topologies remains an untapped frontier. Determining the ideal placement of distributed assemblies to achieve specific scattering control objectives will be essential. This evolution toward self-organizing metasurface swarms will ultimately realize the vision of software-defined EM space, capable of holistic information and energy management across expansive and disordered environments.

## Methods
**Optimization algorithm.**
The optimization of the metasurface configurations is performed using a genetic algorithm (GA), which is naturally suited for this problem owing to the discrete nature of the metasurface states. Each metasurface element is encoded in a binary representation, and a population consists of a set of binary chromosomes describing the collective state of the metasurface cluster. For each individual in the population, the global far-field scattering response is efficiently computed using the nodal network. The resulting scattering field is then evaluated by a predefined fitness function, which quantifies the similarity between the reconstructed scattering pattern and the target response. Based on the fitness values, individuals are ranked and subjected to standard genetic operations, including selection, crossover, and mutation, to generate the next generation. This evolutionary process is iterated until either a predefined maximum number of generations is reached or the fitness function converges beyond a prescribed threshold.

For multi-beam radiation scenarios, the target scattering pattern depends on both the number of beams and the corresponding energy allocation. In this work, we assume equal power distribution among all beams. Accordingly, the fitness function is defined as:

$$fitness = \omega_1 \sum_{i \neq j}(f_i - f_j)^2 + \omega_2 \sum \frac{1}{f_i^2} \tag{3}$$

Where $f_i$ denotes the radiated intensity (realized gain) of the i-th beam in its intended steering direction. The first term penalizes the pairwise imbalance among different beams, enforcing uniform gain distribution across all beams. The second term promotes high overall radiation efficiency by encouraging the maximization of the gain of each beam. The weighting coefficients balance the trade-off between gain uniformity and absolute gain enhancement.

For optical illusions, the objective is to reconstruct the scattered field of the metasurfaces such that it closely matches the scattering signature of a prescribed target object. We first define a shape similarity metric to quantify the resemblance between the reconstructed and target scattering patterns as:

$$R = \frac{\sum_i (f_{rec,i} - \bar{f}_{rec})(f_{tar,i} - \bar{f}_{tar})}{\sqrt{\sum_i (f_{rec,i} - \bar{f}_{rec})^2}\sqrt{\sum_i (f_{tar,i} - \bar{f}_{tar})^2}} \tag{4}$$

Where $f_{rec,i}$ and $f_{tar,i}$ denote the reconstructed and target scattered-field intensity at the i-th angular direction, and $\bar{f}_{rec}$ and $\bar{f}_{tar}$ denote the mean value of the scattering field averaged over all sampled directions. The fitness function is then further expressed as $fitness = 1 - R$.

**Sample fabrication.**
To validate the programmability of EM space enabled by distributed metasurfaces, we fabricate three identical metasurfaces with 8*8 independently programmable meta-atoms, a transmitter antenna array, and three power-divider prototypes using standard printed circuit board (PCB) technology. The corresponding photographs are displayed in Fig. 5a. Each metasurface has an effective aperture of 143.4 mm × 143.4 mm and integrates 64 PIN diodes to enable real-time local impedance reconfiguration. A commercial PIN diode (model: MACOM MADP-000907-14020 PIN Diode) is employed in the design with its equivalent circuit model shown Fig. S6. Eighteen control connectors are distributed along the periphery of each metasurface to enable independent addressing of individual meta-atoms. The power dividers are interfaced via SMA connectors.

**Experimental characterization.**
The excitation antenna array and the reference objects were mounted on a motorized rotator, supported by foam pillar (with a relative permittivity of approximately 1 to minimize interference with the scattered fields). The antenna was connected to a custom-designed equal-split phase-shifting power divider (Fig. S14). By assigning specific reference phases to the eight feeding ports, arbitrary control over the radiation beam was achieved. To demonstrate this programmability, three phase configurations were applied. Taking the first port as the phase reference, the phase distributions were set to (0°, 0°, 0°, 0°, 0°, 0°, 0°, 0°), (0°, 30°, 75°, 125°, 170°, 220°, 260°, -35°), and (0°, -30°, -75°, -125°, -170°, -220°, -260°, 35°). These configurations generate three representative beams (#1–#3) with illumination directions of 0°, 20°, and -20°, respectively. To satisfy the far-field criterion, the receiving antenna was positioned 7 m away at the same elevation as the rotation stage. As the stage rotated, the angle θ between the receiving antenna and the scattered field varied, simulating dynamic probing from different detection viewing angles. The far-field scattering intensities were recorded and normalized to their maximum values.

We placed the spatially distributed metasurfaces on custom-designed foam supports. A multi-channel DC voltage control board supplied bias voltages of 0 V or 2.5 V to individual meta-atoms to actively modulate their phase responses. During the continuous rotation of the rotator, the receiving antenna captured the far-field scattering intensity generated by the metasurface clusters, which was similarly normalized to the maximum value.

## Acknowledgements
The work was sponsored by the National Natural Science Foundation of China (No. 62401015, No. U25A20520), the Natural Science Foundation of Anhui Province (No. 2408085QF182 and No. 2508085QF251).

## Data availability
The executable codes and datasets are available from the corresponding author on reasonable request.

## Author contributions
M.L. conceived the research concept, established the theoretical formulation, designed the experimental framework, and drafted the manuscript. L.S. carried out the experiments and contributed to manuscript preparation. D.L. participated in the experimental work. L.M. and G.D. contributed to the numerical simulations. X.Z., C.Q., and Z.W participated in discussions and provided constructive suggestions. L.S., P.Y., and H.C. supervised the project and served as corresponding authors. All authors discussed the results and reviewed the manuscript.

## Competing interests
The authors declare no competing interests.

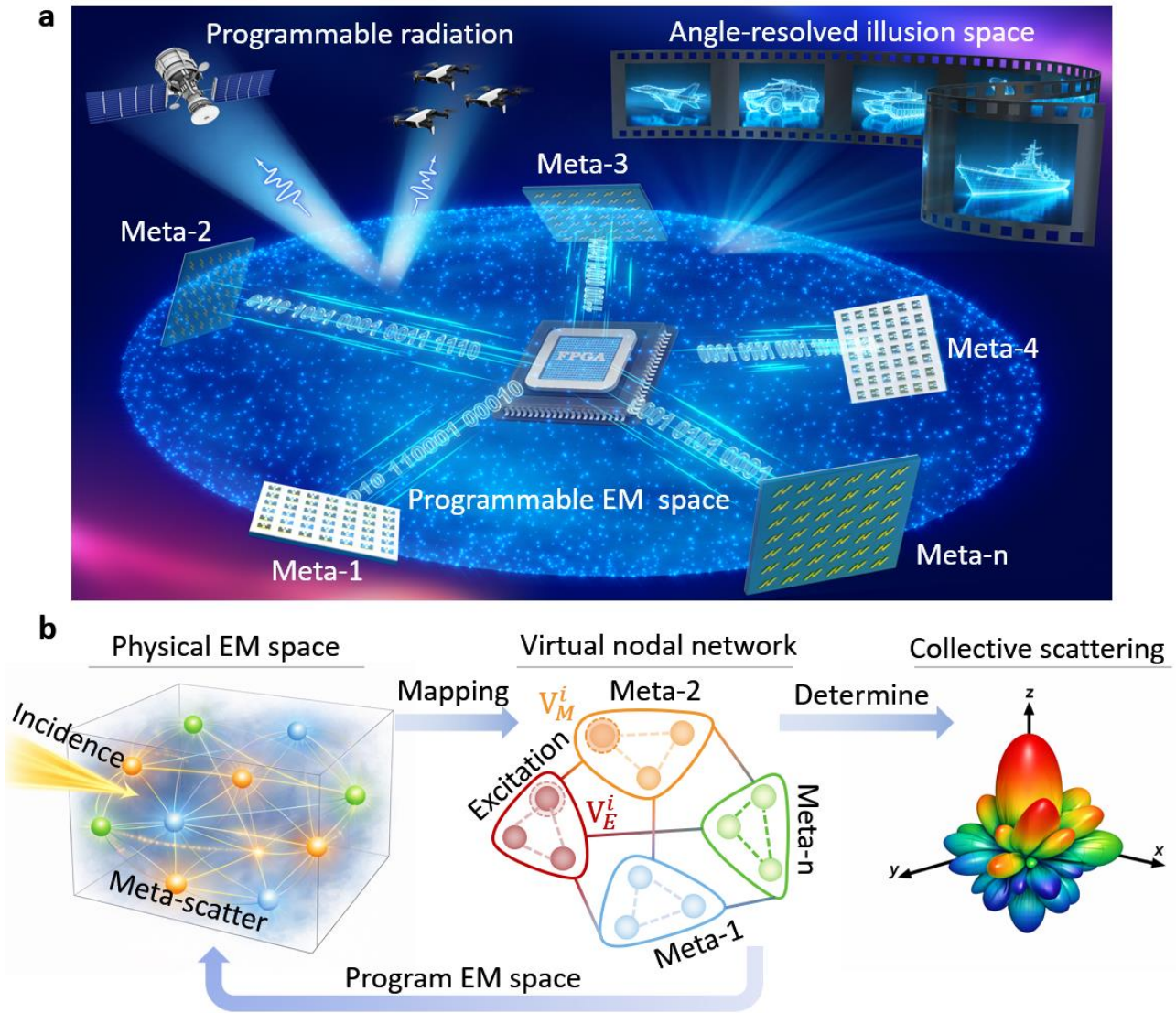

**Fig. 1 | Programmable electromagnetic space enabled by cooperative metasurface clusters. a,** Concept of programmable electromagnetic (EM) space formed by cooperative metasurface clusters. Spatially distributed metasurface (Meta-1 to Meta-n) collectively form an extended EM space whose global scattering can be programmed through coordinated control of individual meta-atoms. By tailoring the collective interactions among distributed metasurfaces, diverse space-level wave functionalities can be realized, including programmable collective radiation and creating angle-resolved illusion spaces. **b,** Deterministic programming framework based on a virtual nodal network. Complex EM field landscapes in physical space are mapped onto a virtual nodal network where excitations and all meta-atoms are represented as interconnected nodes. The nodal states of the network determine the resulting collective scattering response, while inverse programming of the nodal states enables deterministic control of the EM space formed by the metasurface clusters.

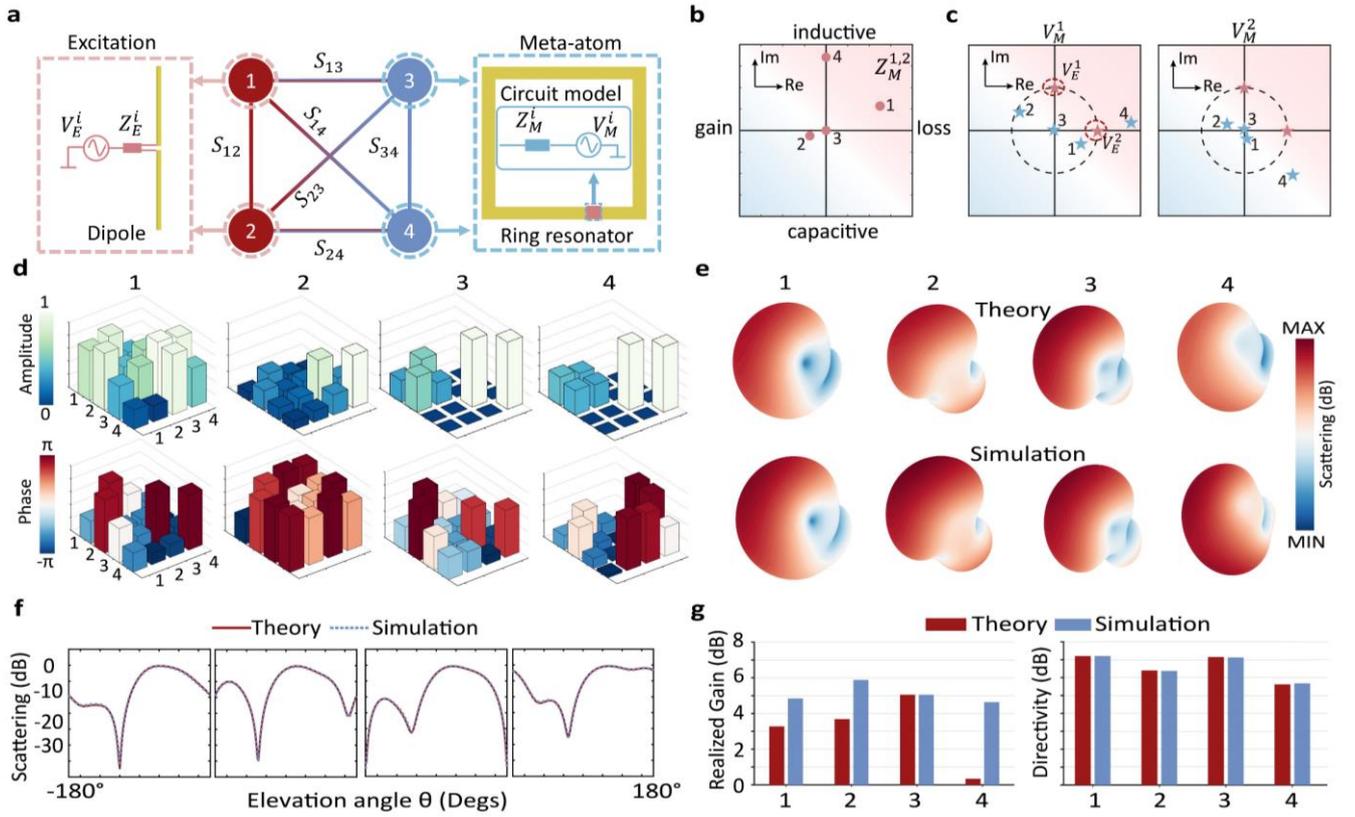

**Fig. 2 | Mapping of physical electromagnetic scenes onto a virtual nodal network. a**, Configuration of a representative EM scene and its mapping onto virtual nodal network. The physical system consists of two dipole antennas acting as excitation sources and two meta-atoms, exemplified by square split-ring resonators. For each meta-atom, an arbitrarily selected deeply subwavelength region (red area) is assigned a virtual lumped port, characterized by a nodal voltage $V_M^i$ and impedance model $Z_M^i$. **b**, Distribution of four representative impedance states ($Z_M^1 = Z_M^2$) on the complex plane. and **c**, Calculated nodal voltage $V_M^1$, $V_M^2$ of the two meta-nodes corresponding to the impedance states shown in **b**, red stars represent the excitation-node voltages. **d**, Scattering matrix for quantifying all-to-all mutual coupling. **e**, Three-dimensional far-field scattering patterns calculated using the network theory and obtained from full-wave simulations. **f,** Two-dimensional far-field scattering patterns, comparing theory (dashed lines) with full-wave simulations (solid lines). **g**, Comparison of realized gain and directivity between the theory and full-wave simulations.

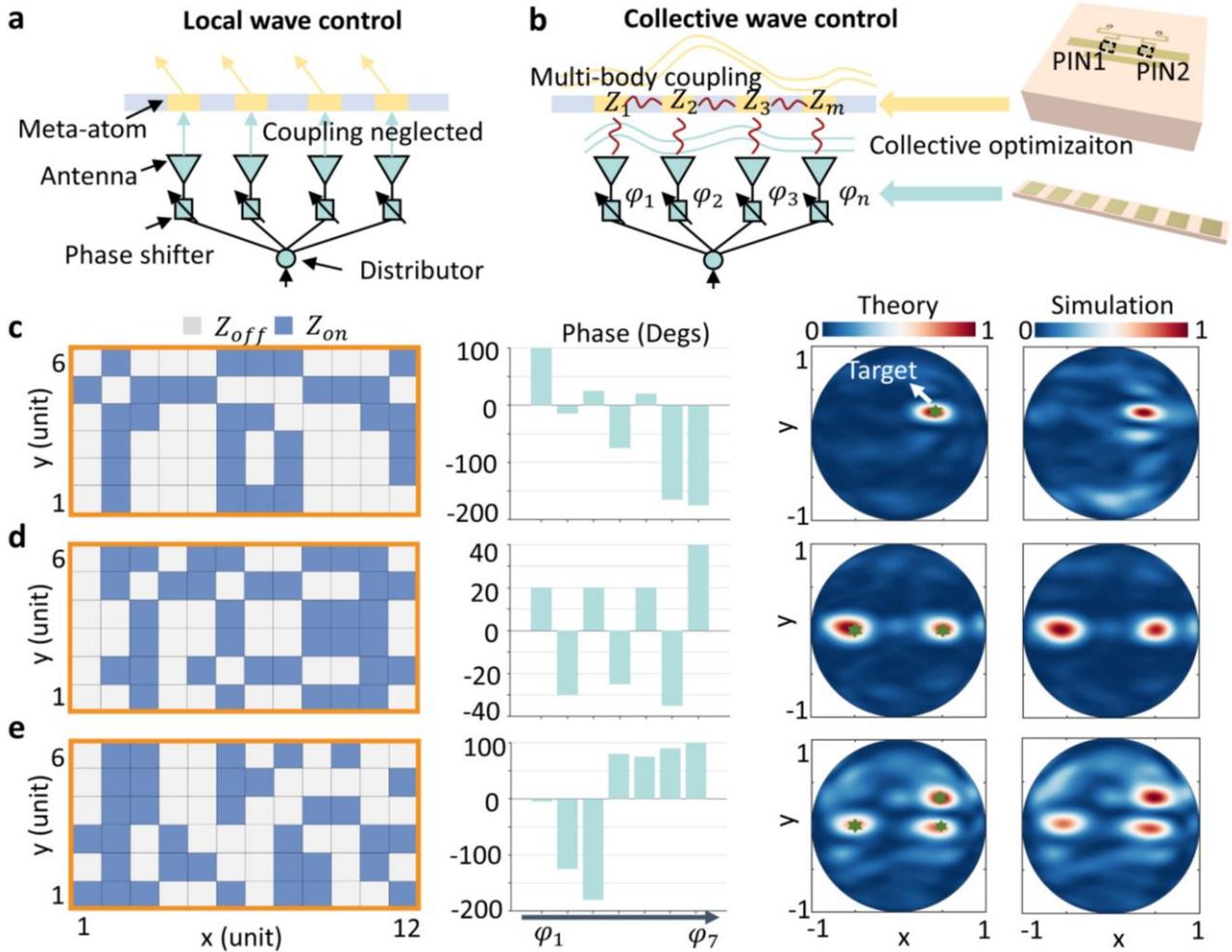

**Fig. 3 | Programmable collective radiation by a deeply coupled meta-emitter. a,** Local wave control with uncoupled meta-atoms. In conventional metasurface designs, inter-element coupling is typically neglected and wave manipulation relies on local phase engineering of isolated elements. **b,** Collective wave control in a deeply coupled metasystem. In the proposed framework, the incident wave and metasurface form a strongly coupled many-body system, where inter-element interactions collectively determine the radiation. Both the meta-atom impedance states ($Z_1$–$Z_m$), controlled through PIN diodes, and the excitation phases ($\phi_1$–$\phi_n$) are jointly optimized to synthesize the desired collective radiation. **c–e,** Programmable collective radiation generated by the deeply coupled meta-emitter. Results are shown for **c** single-beam, **d,** dual-beam, and **e,** triple-beam configurations. For each case: the left panels show the optimized metasurface configurations, the middle panels present the excitation phase distributions, and the right panels compare the radiation patterns predicted by the theory with full-wave simulations. The results demonstrate programmable multifunctional radiation with excellent agreement between theory and simulation.

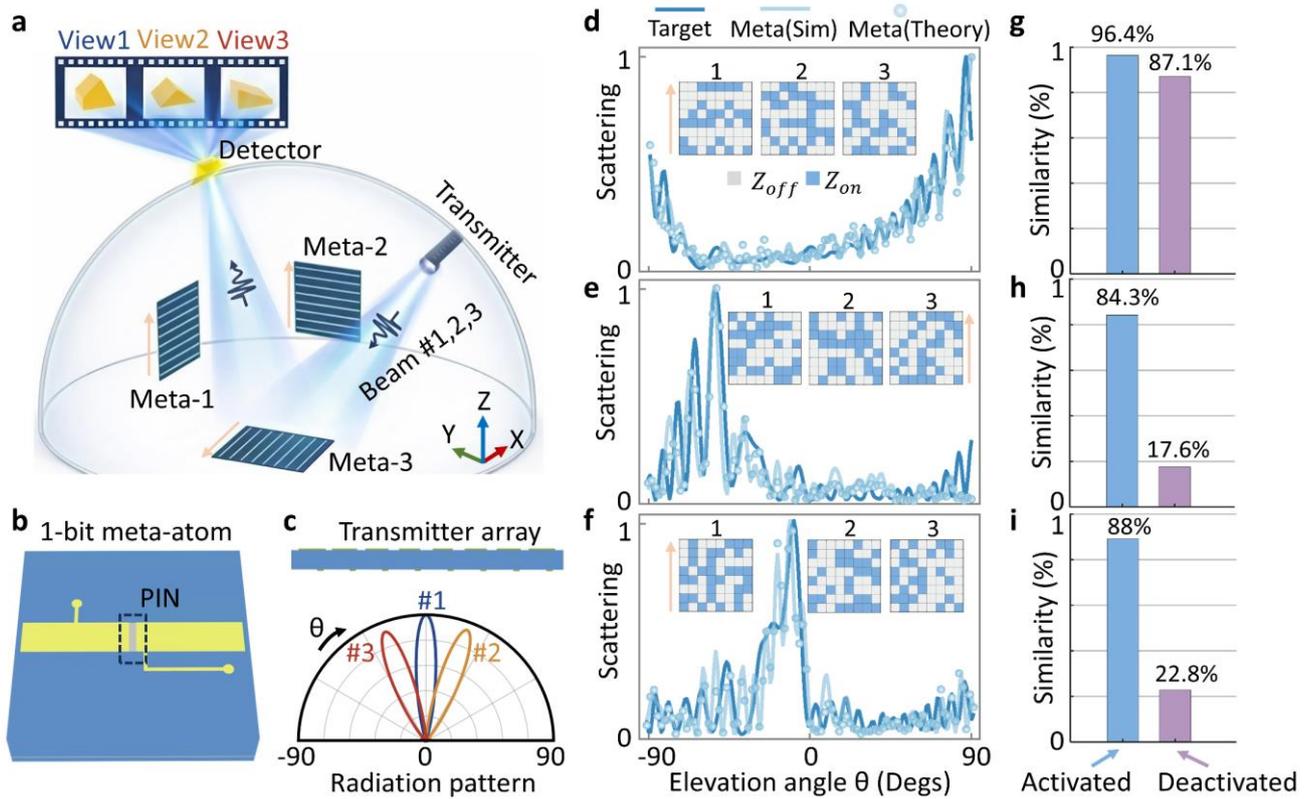

**Fig. 4 | Programmable angle-resolved illusion space created by metasurface clusters. a**, Concept of programmable angle-resolved illusion space, where multiple metasurfaces are spatially distributed in an open space and illuminated by a transmitter array that generates multiple excitation beams. Through cooperative scattering of the metasurface cluster, different far-field scattering landscapes are produced toward different observation directions (View 1–View 3), enabling perspective-dependent optical illusions detected by a remote receiver. **b**, Schematic of the designed 1-bit reflective meta-atom. **c**, The transmitter array generates multiple excitation beams (#1–#3) with different radiation directions, which illuminate the metasurface cluster and trigger distinct collective scattering responses across the EM space. **d–f**, Angle-resolved scattering responses generated by the metasurface cluster. Three representative illusion cases are shown. The dark solid curves represent the target scattering profiles of object 1, 2, and 3; the light solid curves correspond to full-wave simulations of the metasurface-reconstructed scattering; and the scatter points denote the results predicted by theoretical calculation. **g–i**, The similarity between the target scattering distributions and the reconstructed ones is evaluated under metasurface-activated and metasurface-deactivated conditions. High similarity is achieved when the metasurface cluster is activated, whereas the similarity significantly decreases when the metasurfaces are deactivated, confirming that the illusion space originates from the cooperative scattering of the metasurface clusters.

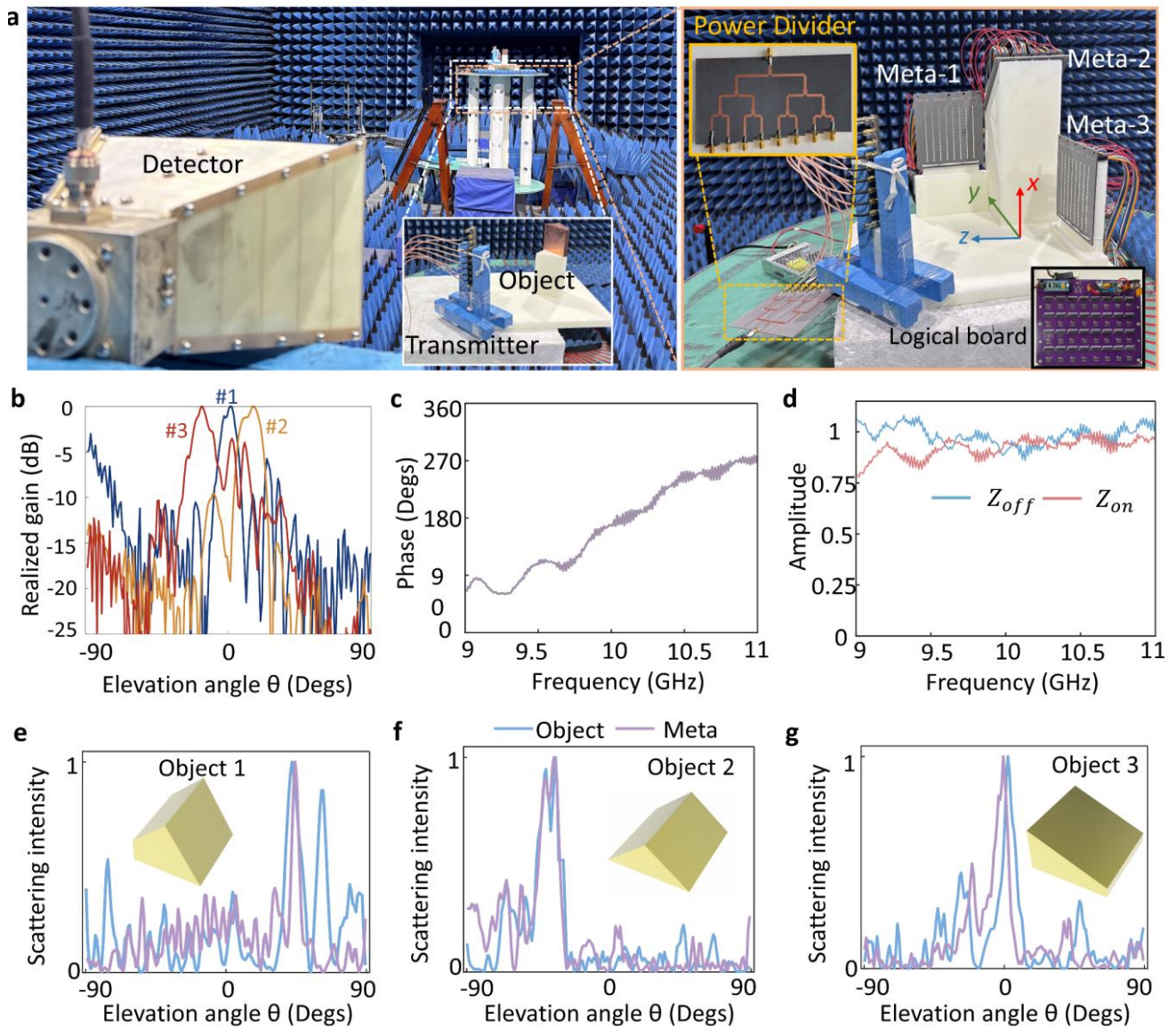

**Fig. 5 | Experimental demonstration of metasurface clusters for programmable illusion space. a,** Schematic of the experimental setup. **b,** Radiation characteristics of the transmitter array. The transmitter generates three excitation beams (#1–#3) with different illumination directions, exciting the EM space formed by the metasurface clusters. **c,** Phase and **d,** Amplitude responses of the programmable metasurface. **e-g,** Programmable illusion under different illumination directions. Panels e, f and g correspond to illumination by beam #1, #2 and #3, respectively. The scattering pattern generated by the metasurface cluster (Meta) closely reproduce those of the corresponding reference objects (Object), demonstrating programmable angle-resolved illusion space.